\documentclass[aps,ams,amsmath,prx,twocolumn]{revtex4-1}
\usepackage{graphics}
\usepackage{float}
\usepackage{graphicx}
\usepackage{epsfig}
\usepackage{amsmath}
\usepackage{amssymb}
\usepackage{amsfonts}
\usepackage{bm}
\usepackage{color}
\usepackage{bbm}
\usepackage{hyperref}
\usepackage[normalem]{ulem}
\usepackage{soul}
\usepackage{tabularx}

\begin{document}

\title{Nonlocal thermoelectric resistance in vortical viscous transport}

\author{Songci Li}
\affiliation{Department of Physics, University of Wisconsin-Madison, Madison, Wisconsin 53706, USA}

\author{Alex Levchenko}
\affiliation{Department of Physics, University of Wisconsin-Madison, Madison, Wisconsin 53706, USA}

\begin{abstract}
The pursuit for clearly identifiable signatures of viscous electron flow in the solid state systems has been a paramount task in the search of the hydrodynamic electron transport behavior. In this work, we investigate theoretically the nonlocal electric and thermal resistances for the generic non-Galilean-invariant electron liquids in the multiterminal Hall-bar devices in the hydrodynamic regime. The role of the device inhomogeneity is carefully addressed in the model of the disorder potential with the long-range correlation radius. We obtain analytic expressions for the thermoelectric resistances that are applicable in the full crossover regime from charge neutrality to high electron density. We show that the vortical component of the electron flow manifests in the thermal transport mode close to the charge neutrality where vorticity is already suppressed by the intrinsic conductivity in the electric current. This behavior can be tested by the high-resolution thermal imaging probes. 
\end{abstract}

\date{June 7, 2022}
\maketitle

\section{Introduction}\label{sec:I}

The electron liquids in solids can exhibit hydrodynamic transport behavior provided sufficient sample purity and proper regime of temperatures \cite{Gurzhi-UFN}. These special conditions are required to ensure that the equilibration length of electron collisions is short compared to the momentum relaxation length scales associated with the scattering off impurities and phonons. In this regime macroscopic description applies and the flow of electrons through a crystal can be described within the limits of hydrodynamic theory. 

For a long time, the key manifestation of electron hydrodynamics in the transport properties of solids was attributed to the prediction made by Gurzhi of negative differential resistivity, $\partial\rho/\partial T<0$, namely the decrease of sample resistance with the increase of temperature \cite{Gurzhi-JETP}. It is noteworthy to mention that this result is counterintuitive at the first glance. Indeed, lattice degrees of freedom are activated with the increase of temperature, electron scattering probability grows, which must lead to an increase of resistivity. However, it can be shown that if electrons attain the hydrodynamic limit via frequent electron-electron collisions, the resistive flow is determined by viscosity with momentum relaxation occurring at the sample boundaries. For Fermi liquids, viscosity decreases with the increase of temperature \cite{AK}; therefore resistivity diminishes as well. This signature behavior was observed in the high-mobility semiconductor quantum wires and two-dimensional quantum wells \cite{Molenkamp,deJong}.    

In recent years, the advent of boron nitride encapsulated graphene renewed interest in the topics of electron hydrodynamics; see reviews \cite{NGMS,Lucas-Fong,Polini-Geim}. This stimulated research activities and led to insightful predictions, many of which were already validated experimentally. For instance, the Dirac fluid in graphene was expected to display anomalous thermoelectric responses most  pronounced near charge neutrality~\cite{Aleiner,Lucas,Principi-2DM,LLA,LAL}. This was confirmed by the observation of gross violation of the Wiedemann-Franz law, which manifests by the significant increase in the Lorenz ratio~\cite{Crossno} and the Seebeck coefficient~\cite{Ghahari} as a function of carrier concentration in the carefully controlled domain of the phase diagram defined by temperature and particle density. The collective character of the viscous flow was predicted to enhance conductance in the electron transport through microconstrictions  as compared to its value in the ballistic limit \cite{Guo}. This effect was clearly demonstrated in graphene devices with engineered quantum point contacts \cite{Kumar} and electrostatic dams defined by lateral $p$-$n$ junction barriers \cite{Brar}. Another peculiar aspect of the hydrodynamic transport is defined by its nonlocality that may result in the formation of current vortices concomitant with the appearance of the negative nonlocal resistances \cite{Falkovich,Levitov}. These features attracted significant theoretical attention \cite{Torre,Pellegrino,Shytov,Xie,Danz} and affirmative experimental tests \cite{Bandurin-1,Bandurin-2}. Furthermore, scanning tunneling and magnetic imaging probes were successfully applied to directly visualize electronic flows in restricted geometries \cite{Sulpizio,Ku,Jenkins,Ilani}. The compelling evidences for the characteristic Poiseuille flow profile were identified in the Hall-bar devices as manifestations of electronic viscous effects. In contrast, the whirlpool character of the electronic flow remained elusive up until very recently when geometrical decoration of the device combined with an elaborate superconducting quantum interference device  (SQUID)--on-tip imaging technique confirmed the existence of current vortices \cite{Zeldov}.  

In this work, we dwell further on the issues related to the emergence of thermoelectric current vortices and the sign-alternating value of the corresponding nonlocal resistances. The focal point of our study concerns the role of Galilean invariance. To set the stage for the topic, we recall the basic fact that in the Galilean-invariant liquids, the particle current is given by the momentum density divided by the particle mass~\cite{LL-V6}. For electron liquids in crystals, this is possible only in the case of a single partially occupied band with a strictly parabolic dispersion, which can be a valid approximation only in some cases. Thus generically, in most physical systems including graphene, the electron liquid does not possess Galilean invariance. Therefore, particle and entropy currents are described not only by the hydrodynamic mode associated with the onset of fluid motion, but also by a relative mode described by a dissipative matrix of intrinsic thermoelectric coefficients. We aim to trace the evolution of the vortical component of viscous electron flow as the system is tuned toward the charge neutrality point where intrinsic dissipative processes dominate over the hydrodynamic mode. We concurrently study the changes in the nonlocal resistances characteristic to the particular flow pattern.   

The rest of this paper is organized as follows. In Sec.~\ref{sec:hydro}, we briefly introduce the main ingredients of the hydrodynamic formalism. In Sec.~\ref{sec:resistances}, we study transport in a long Hall-bar device. We consider geometry with transverse injection of the current in two measurement setups including electrical and thermal biasing scenarios. For each case we derive the corresponding nonlocal resistance in the multiterminal measurement configuration. In Sec.~\ref{sec:summary}, we provide a summary of the main results and concluding remarks. 


\section{Hydrodynamic Formalism}\label{sec:hydro}
The hydrodynamic equations are concerned with the conservation of the electron density $n$, entropy density $s$, and momentum flux of the electron liquid. The first two conservation laws in linear response are expressed by the continuity equation
\begin{equation}
	 \frac{\partial\{n,s\}} {\partial t} + \boldsymbol{\nabla}\cdot {\bf j}_{\{n,s\}}=0.
\end{equation}
The particle current density ${\bf j}_n$ and entropy current density ${\bf j}_s$ can be written as the sum of two contributions: (i) current densities carried by the hydrodynamic velocity ${\bf v}(\mathbf{r})$, and (ii) current densities carried by the transport relative to the electron liquid as driven by the gradients of the electrostatic potential $\phi$ and temperature $T$. Therefore, we express the constitutive relations for ${\bf j}_n$ and ${\bf j}_s$ as follows:
\begin{subequations}
	\begin{align}
		& {\bf j}_n = n{\bf v} -\frac{\sigma}{e^2} e\boldsymbol{\nabla} \phi - \frac{\gamma}{T} \boldsymbol{\nabla} T, \label{eq:jn} \\
		& {\bf j}_s = s{\bf v} -\frac{\gamma}{T} e\boldsymbol{\nabla} \phi - \frac{\kappa}{T} \boldsymbol{\nabla} T, \label{eq:js}
	\end{align}
\end{subequations}
with $\kappa,\, \sigma,\,\gamma$ being the thermal conductivity, the intrinsic conductivity, and the thermoelectric coefficient of the electron liquid, respectively. In the constitutive relations, we do not assume Galilean invariance, wherein $\sigma=0,\,\gamma=0$. In the regime of low doping $n\ll s$, the thermoelectric coefficient may be estimated to scale as $\gamma/T\sim n/s$. In the same limit, the intrinsic conductivity $\sigma$ is nearly a constant modulo logarithmic renormalizations in the weak-coupling theory \cite{Mishchenko,Kashuba}. 

The conservation of momentum density is expressed in terms of the electronic Navier-Stokes (N-S) equation. The latter, in the steady state and the linear response to the applied forces, is written as
\begin{align}\label{eq:NS}
	\eta \nabla^2 {\bf v} -\alpha{\bf v} =ne \boldsymbol{\nabla} \phi + s \boldsymbol{\nabla} T,
\end{align}
where $\eta$ is the shear viscosity. For systems tuned sufficiently close to charge neutrality, it scales as $\eta\propto T^2$ \cite{Fritz}. The terms on the right-hand side are the driving forces of the hydrodynamic flow. The first term on the left-hand side is the viscous stress, while the second term captures the friction force due to disorder. One of the main sources of disorder originates from charge impurities in the substrate on which the graphene flake is deposited \cite{Crommie,LeRoy}. These impurities induce spatial fluctuations in the chemical potential and lead to local regions of positive and negative charge density, commonly referred to as charge puddles. Averaging the flow of the electron fluid over the spatial inhomogeneities leads to an appearance of the effective friction force ${\bf F}_f=-\alpha{\bf v}$, with the friction coefficient $\alpha$ given by \cite{LLA}
\begin{align}\label{eq:alpha}
	\alpha = \frac{\langle \left(s\delta n -n \delta s\right)^2 \rangle}{2\left(\frac{n^2\kappa}{T}-\frac{2ns\gamma}{T}+\frac{s^2\sigma}{e^2}\right)},
\end{align}
where $\delta n({\bf r})$ and $\delta s({\bf r})$ denote local fluctuations of the particle and entropy density, respectively, and brackets $\langle \cdots \rangle$ define spatial averages. In this macroscopic description, namely upon spatial averaging over the distances larger than the correlation radius of the disorder potential, the number density and the entropy density in Eq.~\eqref{eq:NS} can be taken to be spatially uniform. It should be noted though, that averaging also leads to effective renormalizations of both these densities, as well as intrinsic kinetic coefficients. In the treatment that follows, we assumed that all these renormalizations were absorbed into correspondingly redefined quantities. For the bulk samples and Hall-bar devices, this analysis was carried out in the recent work of Refs. \cite{LLA,LAL}. For instance, the renormalized intrinsic conductivity, $\sigma+e^2\chi$, acquires positive correction with $\chi=\frac{1}{2\eta}\frac{\langle U^2\rangle}{(2\pi e^2)^2}$, where $U$ is random potential. This signifies the conductivity enhancement by the correlated effects of electron and disorder scattering. 
 

 \section{Thermoelectric Resistances}\label{sec:resistances}
 
\begin{figure}
    \includegraphics[width=3in]{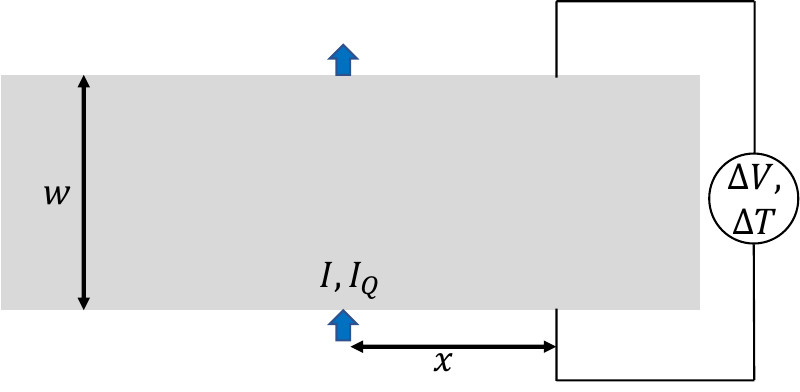}
    \centering
    \caption{The geometric setup: the electric (thermal) current $I(I_Q)$ is injected at $x=0,y=0$ and drained at the opposite end $x=0,y=w$. The nonlocal electrical (thermal) resistance is measured at a distance $x$ away from the injection site.}
    \label{fig:nonlocal}
\end{figure}

In the present paper, we in part revisit the hydrodynamic flow proposed in Ref. \cite{Falkovich}, i.e., the flow in a strip of finite width $w$ ($-\infty < x < \infty,\, 0<y<w$) with source and drain contacts placed at the opposite edges of the strip ($x=0,\, y=0,w$); see Fig.~\ref{fig:nonlocal}. We further extend previous considerations \cite{Falkovich,Levitov,Torre,Pellegrino} to include thermal effects and highlight the difference in the emergence of the vortical effects. 

\subsection{Electrically Biased Setup}
In this section, we consider the flow pattern driven by the transverse current injected and drained through the contacts. Since the system is purely electrically biased, there is no temperature gradient $\nabla T$. The viscous flow is described by the N-S equation and the continuity equation
\begin{align}
	& \nabla^2 {\bf v} - \frac{1}{l^2_G} {\bf v} = \frac{ne}{\eta} \boldsymbol{\nabla} \phi, \label{eq:NS-1} \\
	& \boldsymbol{\nabla} \cdot {\bf j} = 0, \quad {\bf j} = e{\bf j}_n = ne {\bf v} - \sigma \boldsymbol{\nabla} \phi, \label{eq:jn-1}
\end{align}
where $l_G=\sqrt{\eta/\alpha}$ is the generalized Gurzhi length. We recall that the conventional definition of the Gurzhi length is given by the geometrical mean between the momentum-conserving electron mean free path and the momentum-relaxing one on the point-like quenched disorder. Therefore, its temperature dependence is governed by the former length scale. In the present case, the frictional processes are captured by the coefficient $\alpha(n,T)$ that exhibits a rather complex density and temperature dependence as should be clear from Eq.~\eqref{eq:alpha}. Therefore, the Gurzhi crossover from the flow in the restricted geometry to that in the bulk, which is described by the relative importance of the Laplacian and local friction term in Eq.~\eqref{eq:NS}, is more intricate in the present model. 

Introducing the stream function $\psi$ via ${\bf v}={\bf z} \times \boldsymbol{\nabla} \psi = \left(-\partial_y \psi, \partial_x \psi\right)$, it is evident from the N-S equation~\eqref{eq:NS-1} that the stream function satisfies
\begin{align} \label{eq:psi-1}
	\left(\nabla^4 - \frac{1}{l^2_G} \nabla^2\right) \psi=0.
\end{align}
To solve for $\psi$, boundary conditions (b.c.) must be specified. One of the b.c. is the no-slip boundary condition $v_t=0$, where the subscript ``$t$" stands for the tangential component. In terms of the stream function, this is equivalent to $\partial_n \psi=0$, where the subscript ``$n$" stands for the normal component. The second b.c. is imposed on the normal component of the velocity by the injected and drained current $I(\mathbf{r})$ flowing in and out through the contacts. With the constitutive relation in Eq.~\eqref{eq:jn-1}, this reads
\begin{align}
	I (\mathbf{r}) = ne v_n - \sigma \partial_n \phi = ne \left[ \left(1+ \frac{l^2_n}{l^2_G}\right) v_n - l^2_n \nabla^2 v_n\right],
\end{align}
where $l_n = \sqrt{\eta\sigma}/ne$, and we have used the N-S equation~\eqref{eq:NS} to relate the potential gradient $\partial_n \phi$ to the flow velocity $v_n$. 

Similarly to Ref.~\cite{Falkovich}, we express $\psi$ in a mixed coordinate-momentum representation,
\begin{align}\label{eq:Fourierpsik}
	\psi(x,y) = \int \frac{dk}{2\pi} e^{ikx} \psi_k(y),
\end{align}
and find that the function $\psi_k(y)$ satisfies the equation
\begin{align} \label{eq:psik(y)}
	(\partial^2_y-k^2) (\partial^2_y-q^2) \psi_k(y) = 0, \quad q = \sqrt{k^2 + l^{-2}_G},
\end{align}
coupled with the no-slip b.c. 
\begin{align}\label{eq:noslip-1}
	\partial_y \psi_k(y)|_{y=0,w} = 0,
\end{align}
and the b.c. of current injection 
\begin{align}\label{eq:currentinjection-1}
	\frac{I}{ikne} = (1 + \ell) \psi_k(y) |_{y=0,w} - l^2_n (\partial^2_y-k^2) \psi_k(y) |_{y=0,w}.
\end{align}
The dimensionless parameter $\ell$ is defined as $ \ell = l^2_n/l^2_G$ and we assume pointlike contacts $I(x)=I\delta(x)$. The solution of Eq.~\eqref{eq:psik(y)} is straightforwardly written by a sum of four exponents,
\begin{align}\label{eq:psik(y)_1}
	\psi_k(y) = \frac{I}{ikne} \left(a_+ e^{ky} + a_- e^{-ky} + b_+ e^{qy} + b_- e^{-qy}\right).
\end{align}
Combining such a solution with Eqs.~\eqref{eq:noslip-1} and~\eqref{eq:currentinjection-1}, and solving for the expansion coefficients $a_\pm$ and $b_\pm$, we get
\begin{subequations}
\begin{align}
	& a_+\! =\! \frac{\left(e^{qw}-1\right)q}{(1+\ell) q \left(e^{qw}-1\right) \left(e^{kw}+1\right) - k \left(e^{qw}+1\right) \left(e^{kw}-1\right)}, \\
	& a_-\! =\! \frac{e^{kw} \left(e^{qw}-1\right)q}{(1+\ell) q \left(e^{qw}-1\right) \left(e^{kw}+1\right) - k \left(e^{qw}+1\right) \left(e^{kw}-1\right)}, \\
	& b_+\! =\! -\frac{\left(e^{kw}-1\right)k}{(1+\ell) q \left(e^{qw}-1\right) \left(e^{kw}+1\right) - k \left(e^{qw}+1\right) \left(e^{kw}-1\right)}, \\
	& b_- \!=\! -\frac{e^{qw}\left(e^{kw}-1\right)k}{(1+\ell) q \left(e^{qw}-1\right) \left(e^{kw}+1\right) - k \left(e^{qw}+1\right) \left(e^{kw}-1\right)}.
\end{align}
\end{subequations}

The electric potential can be found from the N-S Eq.~\eqref{eq:NS-1}, which gives $\nabla_i \phi = \left( \eta \nabla^2 v_i - \alpha v_i \right)/ne$. Upon substituting the solution Eq.~\eqref{eq:psik(y)_1} into the N-S Eq.~\eqref{eq:NS-1}, we find the potential
\begin{align}
	\phi(x,y) &= -\frac{I}{2\pi\sigma} \int\limits^{+\infty}_{-\infty} \frac{dk}{k} e^{ikx} \nonumber \\
	&\times \frac{\ell \, q \left[e^{ky} - e^{k(w-y)}\right]}{(1+\ell) \, q \left(e^{kw}+1\right) -  k \left(e^{kw}-1\right)\coth\frac{qw}{2}}, \label{eq:phi_1}
\end{align}
and the nonlocal voltage, $\Delta V(x) \equiv \phi(x,y=0)-\phi(x,y=w)$. Consequently, the nonlocal electric resistance, $R_\text{nl}(x) \equiv \Delta V(x)/I$, is 
\begin{align}
	R_\text{nl}(x) = \frac{1}{\pi\sigma} \int\limits^{+\infty}_{-\infty} \frac{dk}{k}  \frac{\ell q e^{ikx}}{(1 + \ell)\, q \coth\frac{kw}{2} - k \coth\frac{qw}{2}}. \label{eq:Rnl}
\end{align}

We consider limiting cases. For the wide Hall-bar strip, $l_G \ll w$, the nonlocal resistance in the region $|x| \lesssim w$ at different densities is given by the following expressions. 

(1) Low density: $l_n \gg l_G$ or $n \ll \sqrt{\alpha\sigma/e^2}$,
\begin{align}
	R_\text{nl} (x) = 
	\begin{cases}
		\frac{2}{\pi \sigma} \ln\left(\frac{2w}{\pi|x|}\right), & |x| \ll l_G, \\
		\frac{2}{\pi \sigma} \frac{l^2_n}{l^2_n+l^2_G} \ln\left(\frac{2w}{\pi|x|}\right), & l_G \ll |x| \ll w.
	\end{cases}
\end{align}

(2) High density: $l_n \ll l_G$ or $n \gg \sqrt{\alpha\sigma/e^2}$,
\begin{align}\label{eq:Rnl-wide-high}
	R_\text{nl} (x) = \begin{cases}
		\frac{2}{\pi \sigma} \ln\left(\frac{2w}{\pi|x|}\right), & |x| \ll l_n, \\
		-\frac{2\eta}{\pi n^2e^2}\frac{2}{x^2}, & l_n \ll |x| \ll l_G, \\
		\frac{2}{\pi \sigma} \frac{l^2_n}{l^2_n+l^2_G} \ln\left(\frac{2w}{\pi|x|}\right), & l_G \ll |x| \ll w.
	\end{cases}
\end{align}

Conversely, for the narrow strip, $l_G \gg w$, the nonlocal resistance for different densities is given by the following expressions. 

(1) Low density: $l_n \gg w$ or $n \ll \sqrt{\eta\sigma/w^2e^2}$,
\begin{align}
	R_\text{nl} (x) \approx 
		\frac{2}{\pi \sigma} \ln\left(\frac{2w}{\pi|x|}\right), \quad |x| \ll w.
\end{align}

(2) High density: $l_n \ll w$ or $n \gg \sqrt{\eta\sigma/w^2e^2}$,
\begin{align}\label{eq:Rnl-narrow-high}
	R_\text{nl} (x) = \begin{cases}
		\frac{2}{\pi \sigma} \ln\left(\frac{2w}{\pi|x|}\right), & |x| \ll l_n, \\
		-\frac{2\eta}{\pi n^2e^2}\frac{2}{x^2}, & l_n \ll |x| \ll w.
	\end{cases}
\end{align}

In the low-density limit, $R_\text{nl}(x)$ is everywhere positive in the region $|x| \ll w$ and formally identical to that in the Ohmic regime. At high density, we reproduce the results obtained in~\cite{Falkovich} in the spatial region $l_n \ll |x| \ll w$, whereas in the region $|x| \ll l_n$, $R_\text{nl}$ is again formally identical to the Ohmic form. This can be understood from the fact that the constitutive relation is ${\bf j} \approx \sigma {\bf E}$ for $|x| <l_n$.

\begin{figure*}
    \includegraphics[width=0.9\textwidth]{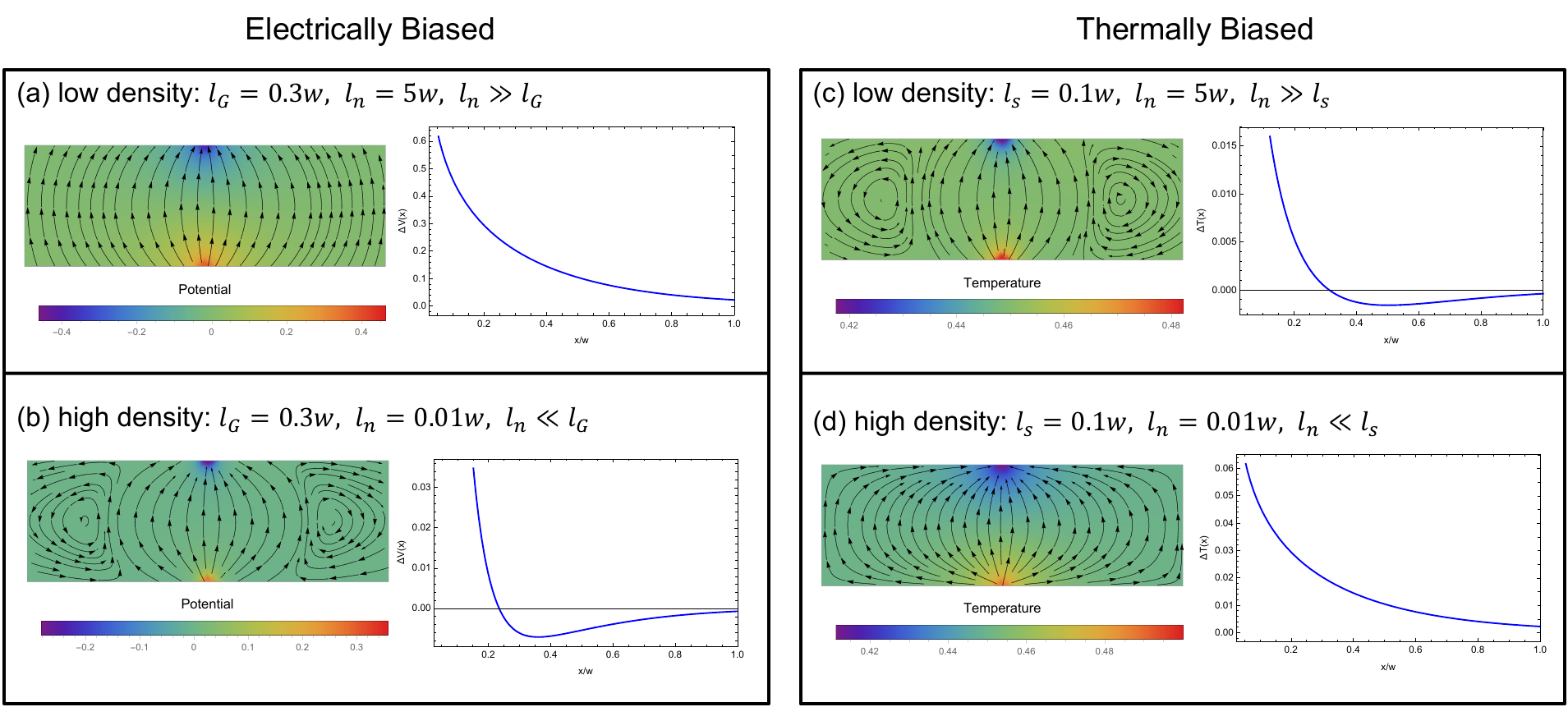}
    \centering
    \caption{A comparative set of plots for the electric and thermal current stream lines at different densities in a Hall-bar device. The stream plot for the flow profile is superimposed on the colored plots for potential or temperature, adjacent to the nonlocal resistance. In each regime the corresponding resistance is computed as a function of distance away from the central line of source-drain contacts. The parameters used in the plots are identical to those in the main text.}
    \label{fig:vortices}
\end{figure*}

\subsection{Thermally Biased Setup}

In this section, we consider the case in which the pair of contacts at $x=0,\,y=0,w$ are kept at different temperatures, so that a transverse entropy current $I_s$ is injected and drained. Since there is no electric current in the bulk, we set ${\bf j}_n$ to zero in Eq.~\eqref{eq:jn} and obtain the relation between the potential, the flow velocity, and temperature gradient:
\begin{align}\label{eq:phi-2}
	e \boldsymbol{\nabla}\phi = \frac{e^2}{\sigma}\left(n{\bf v} - \frac{\gamma}{T} \boldsymbol{\nabla} T\right).
\end{align}
Substituting this relation into the N-S equation~\eqref{eq:NS} and the constitutive relation for ${\bf j}_s$~\eqref{eq:js}, we have
\begin{align}
	& \nabla^2 {\bf v} - \left(\frac{1}{l^2_G} + \frac{1}{l^2_n}\right) {\bf v} = \frac{\varsigma}{\eta} \boldsymbol{\nabla} T, \label{eq:NS-2} \\
	& \boldsymbol{\nabla} \cdot {\bf j}_s = 0, \quad {\bf j}_s = \varsigma {\bf v}- \frac{\varkappa}{T} \boldsymbol{\nabla} T, \label{eq:j-2}
\end{align}
where $\varsigma=s-ne^2\gamma/\sigma T$ and $\varkappa=\kappa- e^2\gamma^2/\sigma T$ are the redefined entropy density and thermal conductivity, respectively. Introducing the stream function $\psi$ in the same mixed coordinate-momentum representation as in Eq.~\eqref{eq:Fourierpsik}, we get
\begin{align} \label{eq:psik(y)_2}
	(\partial^2_y-k^2) (\partial^2_y-q^2) \psi_k(y) = 0, \quad q = \sqrt{k^2 + l^{-2}_G + l^{-2}_n},
\end{align}
coupled with the no-slip b.c.
\begin{align}
	& \partial_y \psi_k(y)|_{y=0,w} = 0, \label{eq:noslip-2}
\end{align}
and the b.c. for thermal current injection,
\begin{align}
	& \frac{I_s}{ik\varsigma} = (1 + \ell) \psi_k(y) |_{y=0,w} - l^2_s (\partial^2_y-k^2) \psi_k(y) |_{y=0,w}, \quad , \label{eq:currentinjection-2}
\end{align}
where $l_s=\sqrt{\eta\varkappa/T\varsigma^2}$. The modified dimensionless parameter $\ell$ acquires the form $\ell = \frac{l^2_s}{l^2_n} \left(1+\frac{l^2_n}{l^2_G}\right)$. In analogy with the electrically biased case, we obtain the temperature distribution 
\begin{align}
	T(x,y) &= -\frac{I_Q}{2\pi\varkappa} \int\limits^{+\infty}_{-\infty} \frac{dk}{k} e^{ikx} \nonumber \\
	&\times \frac{\ell \, q \left[e^{ky} - e^{k(w-y)}\right]}{(1+\ell) \, q \left(e^{kw}+1\right) - k \left(e^{kw}-1\right)\coth\frac{qw}{2}}
\end{align}
and the nonlocal temperature difference, $\Delta T(x) \equiv T(x,y=0)-T(x,y=w)$. Consequently, the nonlocal thermal resistance, $R_\text{th}(x) \equiv \Delta T(x)/I_Q$, is cast in the form
\begin{align}
	R_\text{th}(x) =\frac{1}{\pi\varkappa} \int\limits^{+\infty}_{-\infty} \frac{dk}{k}  \frac{\ell \, q e^{ikx}}{(1+ \ell)\, q \coth \frac{kw}{2} - k \coth \frac{qw}{2}}. 
\end{align}

For simplicity, we consider analytically the limiting cases for the nonlocal thermal resistance at zero friction $\alpha=0$ at different densities. We find for 

(1) the low density: $l_s \ll w \ll l_n$,
\begin{align}\label{eq:Rth-low}
	R_\text{th} \approx 
	\begin{cases}
		\frac{2}{\pi\varkappa} \ln\left(\frac{2w}{\pi|x|}\right), & |x| \ll l_s, \\
		-\frac{2\eta}{\pi T\varsigma^2}\frac{2}{x^2}, & l_s \ll |x| \ll w;
	\end{cases}
\end{align}

(2) the intermediate density: $l_s \ll l_n \ll w$,
\begin{align}
	R_\text{th} \approx 
	\begin{cases}
		\frac{2}{\pi\varkappa} \ln\left(\frac{2w}{\pi|x|}\right), & |x| \ll l_s, \\
		-\frac{2\eta}{\pi T\varsigma^2}\frac{2}{x^2}, & l_s \ll |x| \ll l_n, \\
		\frac{2}{\pi\varkappa} \frac{l^2_s}{l^2_n} \ln\left(\frac{2w}{\pi|x|}\right), & l_n \ll |x| \ll w;
	\end{cases}
\end{align}

(3) High density: $l_n \ll l_s \ll w$,
\begin{align}
	R_\text{th} \approx \frac{2}{\pi\varkappa} \ln\left(\frac{2w}{\pi|x|}\right), \quad |x| \ll w.
\end{align}

The nonlocal thermal resistance is everywhere negative (viscous flow) in the low-density limit and everywhere positive (Ohmic flow) for high density. In the regime sufficiently close to neutrality $\eta\sim (T/v)^2$ and $\varsigma\approx s\sim (T/v)^2$ since $ne^2\gamma/\sigma sT\sim (e^2/\sigma)(n/s)^2\ll1$. Therefore from Eq. \eqref{eq:Rth-low} we conclude that $|R_\text{th}|\propto 1/T^3$ at $x\gg l_s$. The growth of the resistance, in absolute value, with lowering of the temperature is limited by the applicability condition of the hydrodynamic regime. Coincidentally, the temperature dependence of $R_{\text{th}}$ is the same as in Kapitza thermal boundary resistance, $R_{\text{K}}=A/T^3$, although the physical mechanism is completely different \cite{Swartz-Pohl}. Finally, it is worthwhile to note that the nonlocal response of the injected thermal current can also be detected by electric probes. Indeed, the built-in electric potential, which is given by Eq.~\eqref{eq:phi-2}, leads to a nonlocal voltage difference
\begin{align}
	\Delta V(x) = -\frac{e^2\gamma}{\sigma T} \frac{\Delta T(x)}{e} .
\end{align} 
This can be experimentally probed by the scanning tunneling potentiometry (STP) technique; e.g., see Ref. \cite{Brar}.  

\section{Summary and Conclusions}\label{sec:summary}

In this work we consider thermoelectric transport in graphene Hall-bar devices in the hydrodynamic regime. The most peculiar aspect of electron hydrodynamics in graphene is that at charge neutrality the hydrodynamic flow corresponds to a pure heat transport as it carries no charge. The decoupling of charge and heat fluxes was previously explored primarily in light of the observed anomalously large Lorenz ratio. In this study, we provide a comparative analysis for the thermal and electrical transport in the geometry that enables current vorticity. In particular, we trace density evolution of the vortical flow and the concomitant nonlocal resistance in both charge and heat transport modes. In addition, we study how this flow changes across the Gurzhi crossover, namely for narrow and wide samples as compared to $l_G$.    

The main result of our findings is summarized in Fig. \ref{fig:vortices}. We can see that in the charge mode viscous shear flow generates vorticity and a backflow on the side of the main current path in the high-density regime. As a result, the region of negative nonlocal resistance is clear once the measuring contacts are placed in the region of the counterflow. The resistance is proportional to the electron fluid viscosity and inversely proportional to the square of particle density [see Eqs.~\eqref{eq:Rnl-wide-high} and~\eqref{eq:Rnl-narrow-high} for wide and narrow strips, respectively]. This result is in agreement with the previous conclusions presented in Ref.~\cite{Falkovich}, albeit we provide an extension for the Hall-bar devices with long-range disorder. When the density is varied toward the charge neutrality, this vortical picture gradually disappears. In the regime of vanishing carrier doping, the transport is dominated by the relative mode with the finite intrinsic conductivity $\sigma$, and the overall current profile resembles that in the Ohmic limit. The two extreme cases are depicted in panels (a) and (b) of Fig.~\ref{fig:vortices}. In contrast, the thermal transport picture evolves with the opposite trend. Heat current whirlpools become more pronounced close to the charge neutrality where thermal current is dominated by the convective part of the flow, $s\mathbf{v}$, as depicted in panels (c) and (d) of Fig.~\ref{fig:vortices}. The resulting negative thermal resistance falls off quadratically with the distance away from the current injection and also scales inversely proportional to the cube of temperature Eq.~\eqref{eq:Rth-low}. The temperature dependence originates from both electron viscosity and entropy density. 

We highlight that high-resolution thermal imaging and scanning gate microscopes \cite{Zeldov-Nature,Zeldov-Science} and Johnson-Nyquist nonlocal noise thermometry \cite{Skinner,Waissman} may provide exquisite tools to probe viscous electron vorticity in parameter domains where these signatures are no longer present in the charge transport mode.   

\section*{Acknowledgments} 

We gratefully acknowledge insightful discussions with A. Andreev, V. Brar, S. Ilani, P. Kim, Z. Krebs, A. Talanov, and J. Waissman that stimulated this work. This research was financially supported by National Science Foundation Grants No. DMR-1653661 (S.L.) and No. DMR-2203411 (A.L.). This project was conceptualized during the workshop ``From Chaos to Hydrodynamics in Quantum Matter" at the Aspen Center for Physics, which is supported by National Science Foundation Grant No. PHY-1607611.

\end{document}